\documentclass[prl,twocolumn]{revtex4}
\usepackage{graphicx}
\usepackage{amsfonts}

\begin{document}

\title{Diversity-induced resonance in a system of globally coupled linear oscillators}
\author{Ra\'ul Toral$^{1}$, Emilio Hern\'andez-Garc{\'\i}a$^{1}$, James D. Gunton$^{2}$}
\affiliation{1- IFISC (Instituto de F{\'\i}sica Interdisciplinar y Sistemas Complejos),
UIB-CSIC, Campus UIB, E-07122 Palma de Mallorca, Spain \\
2-Department of Physics, Lehigh University, Bethlehem PA 18015, USA
}

\begin{abstract}
The purpose of this paper to analyze in some detail the arguably simplest case of diversity-induced reseonance: that of a system of globally-coupled linear oscillators subjected to a periodic forcing. Diversity appears as the parameters characterizing each oscillator, namely its mass, internal frequency and damping coefficient are drawn from a probability distribution. The main ingredients for the diversity-induced-resonance phenomenon are present in this system as the oscillators display a variability in the individual responses but are induced, by the coupling, to synchronize their responses.  A steady state solution for this model is obtained.  We also determine the conditions under which it is possible to find a resonance effect.  

\end{abstract}
\maketitle
\section{Introduction}

Many situations of interest can be modeled by a forced system whose output depends on some external forcing. Examples can be found in physics, ecology, chemistry, economics, biology and other sciences. For instance, the response of a neuron depends strongly on the input currents, such that a pulse is produced only when the total input surpasses some threshold value. In some cases, the system is under the action of a periodic forcing and the response can be quantified by the amplitude of the periodic oscillations that the system performs at the frequency of the input. A classic example is that of a forced linear pendulum with damping terms. It is known that the response is optimal -the amplitude of the pendulum oscilations reaches a maximum- when the natural frequency of the oscillator matches that of the external forcing. In the case of non-linear oscillators a more complex behavior can appear and the region in which the system synchronizes to the external forcing might depend on the values of the parameters characterizing the system\cite{PRK:2001}.  

A few decades ago, a particularly interesting resonance effect induced by random terms in the dynamics  was discovered. This important phenomenon, known as Stochastic Resonance\cite{BSV:1981,NN:1981}, shows that disorder, in the form of dynamical noise, can improve the response of a non-linear dynamical system to an external stimulus. This counterintuitive effect of noise, first proposed to explain the observed periodicity of ice-ages, has since been extended to a large variety of systems, including bistable, excitable, chaotic systems\cite{GHJ:1998,BHM:1993}. The main result is that, under some very general conditions,  the right amount of noise can help achieve a maximal synchronization of the dynamical variables in response to an external forcing. This phenomenon results in a resonance curve where some adequate measure $R$ of the system's response shows a maximum for a particular value of the noise intensity $D$, with the response diminishing for smaller or larger $D$. The generic mechanism for the resonance is the matching of some characteristic time scale of the forcing (e.g. the period) with a characteristic time induced by the noise (e.g. the Kramers time for jumping over a potential barrier). Examples of this phenomenon occur in a wide variety of fields, including paleoclimatology, chemical reactions, neuronal systems, lasers and biological environments.
 
A different mechanism that also profits from disorder to enhance the response to an external stimulus has been shown in reference \cite{TMTG:2006}. The basic idea is to take advantage of the diversity that appears in an extended system composed of many constituents. It is obvious that in many systems in biology, physics and economics, it is necessary to take into account the fact that the underlying constituents are not identical. The usual assumption that the constituents are identical is only an approximation which is usually taken for reasons of mathematical simplicity. This diversity of the elements can have many different origins: heterogeneity in a parameter (equivalent to quenched noise), heterogeneity in the network of connectivities or in their strength, etc. In fact, one of the main points in \cite{TMTG:2006} is to argue that any source of disorder is able, under the right circumstances, to enhance the response to the forcing. 

More precisely, the authors of reference \cite{TMTG:2006} considered a system composed of $N$ coupled bistable -double well potential- units, subjected to an external weak periodic signal. Diversity is introduced as the variability of a set of parameters $a_k, k=1,\dots,N$ that control the relative stability of each bistable state of the potential.  If the parameter $a_k$ is equal to zero for unit $k$ the two potential wells have the same depth. For $a_k>0$ the well in the right is deeper than the one on the left and vice-versa for $a_k<0$. A weak periodic signal acts upon all the units. By {\sl weak} it is meant that the signal can not induce jumps between the two potential wells in the symmetric case $a_k=0$. It is assumed that the $a_k$'s follow a distribution of average value $\langle a_k\rangle=0$ and variance $\sigma^2$.  It is further assumed that the units are positively  coupled and, hence,  they all tend to stay in the same well. The mechanism for resonance is as follows: If $\sigma=0$, the signal is subthreshold for all units and the system responds globally with just small oscillations; as $\sigma$ increases some of the units are able to follow the signal during the half-period in which the signal goes in the direction of the preferred well;  in the other half of the period,  a different set of units  follows the signal. The units which follow the signal pull the other units, to whom they are attractively coupled, and the collective effect is that a significant fraction of the units is able to  respond to the external forcing. If the diversity $\sigma$ is too large, the units are very different from each other and some of them offer a strong resistance against following the signal. Resistance which can not be overcome by the coupling between units. 

Complementary to this {\sl microscopic} mechanism for resonance in which one analyzes the behavior of individual units, it is also possible to give a {\sl macroscopic} explanation in which one focuses only on the behavior of the global variable that represents the system (usually, simply the sum of all individual variables). The diversity induces a degradation in the global order that results in the lowering of the effective potential barrier separating the two stable states of the collective system. The barrier can then be more easily overcome by the external forcing\cite{TTV:2007}. The collective effect can then be understood as the result of the degradation of order induced by diversity. What is worth noticing is that any source of disorder would lead to a similar effect. This explanation sheds new light on the phenomenon of stochastic resonance in extended systems\cite{LMD:1995,Wio:1996} in which the disorder is induced by noise. In other cases, it is the disorder induced by the competitions in the network of connectivities between the units that induces disorder and drives the resonance\cite{MTS:2008}. As argued in reference \cite{TMTG:2006} the proposed mechanism is generic and might appear in a large variety of systems of interest in many different fields. For instance, the role of the heterogeneous complex network topology in the amplification of external signals has also been addressed in \cite{ALA:2007} and a recent work \cite{CZL:2007} has shown that structural diversity enhances the cellular ability to detect extracellular weak signals. The interplay between noise and diversity in an ensemble of coupled bistable FitzHugh-Nagumo elements subject to weak signal has been considered in \cite{GGK:2007}. Reference \cite{TSCT:2007} reveals that the general mechanism for collective synchronized firing in excitable systems arises from the degradation of order. The ability of diversity to enhance coherent behavior in networks with attractive and repulsive excitable systems has been addressed in \cite{Car:2000,LSAS:2006,TZT:2008} and the combined effects of noise and variability in the synchronization of neural elements has been studied in \cite{GGK:2008}. 

Although surprising at first, the idea that diversity in the units forming a large system can improve the response to an external signal is not against our experience\cite{Pag:2008}. For example, a society can respond to changes in the economy if it is formed by diverse agents, such that a fraction of the population is successful at different times. It is the positive interaction between the agents that can cause  the whole population to respond successfully to the changing environment.

The mechanism proposed for resonance is very simple and requires only generic ingredients. It is the purpose of this paper to analyze in some detail the arguably simplest case: that of a system of globally-coupled linear oscillators subjected to a periodic forcing. Diversity appears as the parameters characterizing each oscillator, namely its mass, internal frequency and damping coefficient are drawn from a probability distribution. The main ingredients for the diversity-induced resonance phenomenon are present in this system as the oscillators display a variability in the individual responses but are induced, by the coupling, to synchronize their responses. The paper is organized as follows: in the next section \ref{sec:model} we will define precisely the model and parameters and solve the equations of motion for the steady state. Once the main formulas are established, in section \ref{results} we will analyze under which conditions it is possible to find the resonance-effect. Finally, in section \ref{conclusions} we will end with a brief conclusion and outlook.

\section{Linear model and solution}
\label{sec:model}
We consider a system of $N$ globally coupled linear oscillators with canonical variables ($x_k,p_k)$, $k=1,\dots,N$, whose evolution is given by:
\begin{equation}
\left\{
\begin{array}{rcl}
\dot x_k & = & \displaystyle \frac{p_k}{m_k}\\
\dot p_k & = & \displaystyle -\frac{\gamma_k}{m_k} p_k-m_k\omega_k^2x_k+\frac{\kappa}{N}\sum_{j=1}^N(x_j-x_k)+F\cos(\Omega t),
\end{array}
\right.
\end{equation}
or the single equivalent equation:
\begin{equation}
m_k\ddot x_k=-\gamma_k\dot x_k -m_k\omega_k^2x_k+\frac{\kappa}{N}\sum_{j=1}^N(x_j-x_k)+F\cos(\Omega t).
\end{equation}
Here, $m_k$, $\gamma_k$ and $\omega_k$ are, respectively, the mass, the damping coefficient and the natural frequency of oscillator $k$. $m_k$ and $\gamma_k$ are assumed to take only positive values. The external forcing $F\cos(\Omega t)$ acts upon all oscillators. The all-to-all coupling term represents the tendency of all oscillators to act synchronously when the coupling coefficient $\kappa$ is positive. This term can also be written as:
\begin{equation}
\frac{\kappa}{N}\sum_{j=1}^N(x_j-x_k)=\kappa(\bar x-x_k),
\end{equation}
where
\begin{equation}
\bar x(t)=\frac{1}{N}\sum_{j=1}^N x_j(t)
\end{equation}
is the collective (or mean-field) variable.
In the absence of forcing, $F=0$, the dynamical system can be written as relaxational dynamics\cite{ST:1999} in a Hamiltonian
\begin{equation}
{\cal H}=\sum_{k=1}^N \left[\frac{p_k^2}{2m_k}+\frac{1}{2}m_k\omega_k^2x_k^2+\frac{\kappa}{2N}\sum_{j=1}^N(x_j-x_k)^2\right],
\end{equation}
in the form
\begin{equation}
\left(\begin{array}{cc} \dot x_k\\ \dot p_k\end{array}\right)
= 
\left(\begin{array}{cc} 0&1\\ -1 & -\gamma_k\end{array}\right)
\pmatrix{\frac{\partial {\cal H}}{\partial x_k}\cr{\frac{\partial {\cal H}}{\partial p_k}}}.
\end{equation}
Since the matrix $\pmatrix{0&1\cr-1&-\gamma_k}$ can be split in symmetric and antisymmetric matrices $\pmatrix{0&0\cr 0 &-\gamma_k}+\pmatrix{0&1\cr-1&0}$, and the symmetric part does not have positive eigenvalues, the Hamiltonian ${\cal H}$ is a Lyapunov function of the dynamics\cite{ST:1999}. It is easy to show directly that the energy decreases steadily:
\begin{equation}
\frac{d{\cal H}}{dt}=\sum_{k=1}^N\left[\frac{\partial {\cal H}}{\partial p_k}\dot p_k+\frac{\partial {\cal H}}{\partial x_k}\dot x_k\right]=-\sum_{k=1}^N\frac{\gamma_k}{m_k}p_k^2\le 0,
\end{equation}
hence the system relaxes to the minimum of ${\cal H}$: $p_k=x_k=0,\forall k$. The external forcing $F\cos(\Omega t)$ injects energy and produces a final state in which the variables $(x_k,p_k)$ oscillate in time with the frequency $\Omega$. Introducing the complex notation $z_k=x_k+i\Omega^{-1}\dot x_k$ we get:
\begin{equation}
m_k\ddot z_k=-\gamma_k\dot z_k -m_k\omega_k^2z_k+\kappa(\bar z-z_k)+F{\rm e}^{-i\Omega t}
\end{equation}
with $\bar z=N^{-1}\sum_{k=1}^Nz_k$. After a transient time, the variables $z_k(t)$ tend to $z_k(t)=a_k{\rm e}^{-i\Omega t}$ with $a_k=|a_k|e^{i\phi_k}$, a set of (complex) constants obtained as:
\begin{equation}
a_k=\frac{F+\kappa a}{m_k(\omega_k^2-\Omega^2)+\kappa-i\gamma_k\Omega}
\end{equation}
Those values must satisfy the self-consistent relation $a=N^{-1}\sum_{k=1}^Na_k$. Straightforward algebra leads to:
\begin{equation}
\label{eq:ak}
a_k=\frac{a}{G}G_k,\hspace{2.0cm} a=\frac{F G}{1-\kappa G},
\end{equation}
where $G_k$  and $G$ are defined as:
\begin{equation}
G=\frac{1}{N}\sum_{k=1}^NG_k,\hspace{1.0cm}G_k=\frac{1}{m_k(\omega_k^2-\Omega^2)+\kappa-i\gamma_k\Omega}.
\end{equation}

In the previous equations, note that coupling acts in two different ways: in the $k$-independent prefactor, i.e. as a collective effect acting equally on all the oscillators, and as the $k$-dependent second factor, which can be understood as a renormalization of the resonant frequency of the oscillator $k$ from $\omega_k$ to $\tilde\omega_k$, with $m_k\tilde\omega_k^2=m_k\omega_k^2+\kappa$.

A standard measure\cite{GHJ:1998} of the global response $R$ is the modulus of the mean variable $\bar z$ normalized to the external input $\displaystyle R=\frac{|\bar z|^2}{F^2}$, or
\begin{equation}
\label{eq:R}
R=\left|\frac{G}{1-\kappa G}\right|^2.
\end{equation}
This is more easily computed in terms of the real and imaginary parts of $G_k=A_k+iB_k$, as:
\begin{equation}
R=\frac{\langle A_k\rangle^2+\langle B_k \rangle^2}{1-2\kappa \langle A_k\rangle+\kappa^2\left(\langle A_k\rangle^2+\langle B_k \rangle^2\right)}
\end{equation}
with:
\begin{equation}
\langle A_k \rangle =\frac{1}{N}\sum_{k=1}^NA_k,\hspace{0.5cm} A_k=\frac{m_k(\omega_k^2-\Omega^2)+\kappa}{\left(m_k(\omega_k^2-\Omega^2)+\kappa\right)^2+\gamma_k^2\Omega^2},
\end{equation}
\begin{equation}
\langle B_k \rangle=\frac{1}{N}\sum_{k=1}^NB_k,\hspace{0.5cm} B_k=\frac{\gamma_k\Omega}{\left(m_k(\omega_k^2-\Omega^2)+\kappa\right)^2+\gamma_k^2\Omega^2}.
\end{equation}

The expressions above are valid whatever the value of the number of oscillators $N$. In the case that the number of oscillators is very large, the sums can be replaced by averages with respect to a probability distribution $f(\Gamma)$ in the space of parameters $\Gamma_k=(m_k,\omega_k,\gamma_k)$:
\begin{equation}
\frac{1}{N}\sum_{k=1} S(\Gamma_k)\rightarrow \langle S(\Gamma_k)\rangle =\int d\Gamma S(\Gamma) f(\Gamma),
\end{equation}
being $f(\Gamma)$ the probability density function of the set of parameters $\Gamma_k$ and $S(\Gamma)$ any function depending on those parameters.

After a transient time, the $k=1,\dots,N$ linear oscillators $z_k=|a_k|e^{i(-\Omega t+\phi_k)}$ oscillate in time with a frequency $\Omega$ and phase $\phi_k$. If all parameters $(m_k,\omega_k,\gamma_k)$ are identical, the modulus $|a_k|$ and the phases $\phi_k$ are identical for all oscillators. If any of the parameters in the set $(m_k,\omega_k,\gamma_k)$ varies from one oscillator to another, there is a dispersion in amplitudes and phases, see figure \ref{fig:6}.

There are several possible ways to quantify this dispersion in the individual dynamical output. A convenient one is to use the variance of the variables $z_k$ normalized by the average value of the modulus squared:
\begin{equation}
\sigma^2_n[z_k]=\frac{N^{-1}\sum_{k=1}^N |z_k-\bar z|^2}{N^{-1}\sum_{k=1}^N |z_k|^2}=\frac{\langle |z_k-\bar z|^2\rangle}{\langle |z_k|^2\rangle}.
\end{equation}
In the case of no dispersion,  $\sigma^2_n[z_k]=0$;  because of the normalization, the maximum value is $\sigma^2_n[z_k]=1$. We define a measure of order as:
\begin{equation}
\label{eq:rho}
\rho=\sqrt{1-\sigma^2_n[z_k]}=\frac{ |\bar z|}{\sqrt{\langle |z_k|^2\rangle}}.
\end{equation}
The definition is such that $\rho=1$ in the case of no dispersion, $\sigma^2_n[z_k]=0$ and decreases to $\rho=0$ as the dispersion increases. Moreover, in the case that all units oscillate with the same amplitude $|z_k|$, the definition reduces to the famous order parameter Kuramoto introduced in his studies of phase synchronization\cite{Kur:1984}. Within the context of this work we will denote by $\rho_{\phi}$ the Kuramoto order parameter:
\begin{equation}
\label{eq:rhophi}
\rho_{\phi}=\left|\frac{1}{N}\sum_{k=1}^Ne^{i\phi_k}\right|=|\langle e^{i\phi_k}\rangle|
\end{equation}
We also introduce a measure of the dispersion in the amplitude of the oscillators, $\rho_{|z|}$ as:
\begin{equation}
\label{eq:rhoz}
\rho_{|z|}=\frac{ \langle |z_k|\rangle }{\sqrt{\langle |z_k|^2\rangle}}.
\end{equation}
It reaches the maximum value $\rho_{|z|}=1$ in the case of no dispersion in the modulus of the oscillators and decreases towards zero as the dispersion increases.

The measure of order $\rho$ can be computed as:
\begin{equation}
\rho=\sqrt{\frac{\langle A_k\rangle^2+\langle B_k \rangle^2}{\langle A_k^2\rangle +\langle B_k^2\rangle }};
\end{equation}
the Kuramoto order parameter can be computed as:
\begin{equation}
\rho_{\phi}=\sqrt{\langle \alpha_k\rangle^2+\langle\beta_k\rangle^2},
\end{equation}
with
\begin{eqnarray}
\alpha_k& =& \frac{m_k(\omega_k^2-\Omega^2)+\kappa}{\sqrt{\left(m_k(\omega_k^2-\Omega^2)+\kappa\right)^2+\gamma_k^2\Omega^2}},\\
\beta_k& =& \frac{\gamma_k\Omega}{\sqrt{\left(m_k(\omega_k^2-\Omega^2)+\kappa\right)^2+\gamma_k^2\Omega^2}};
\end{eqnarray}
and, finally, the measure in the dispersion of the modulus of the oscillators can be written as:
\begin{equation}
\rho_{|z|}=\frac{\left\langle\sqrt{A_k^2+B_k^2}\right\rangle}{\sqrt{\langle A_k^2\rangle +\langle B_k^2\rangle}}.
\end{equation}

\section{Results}
\label{results}

Most of the results in this section (except in one case at the end of the section) consider that the diversity only appears in the natural frequencies $\omega_k$ of the oscillators. These are drawn from a Gaussian distribution of mean $\omega_0$ and variance $\sigma^2$. Since $\sigma$ is a measure of the dispersion in the frequencies, it will be considered as the measure of the diversity. For most of the section, and in order not to introduce too many variable factors in the analysis, we will take the masses, $m_k$, and damping constants, $\gamma_k$, to have the same values in all the units.

\begin{figure}[h]
    \begin{center}
    \includegraphics[width=8cm]{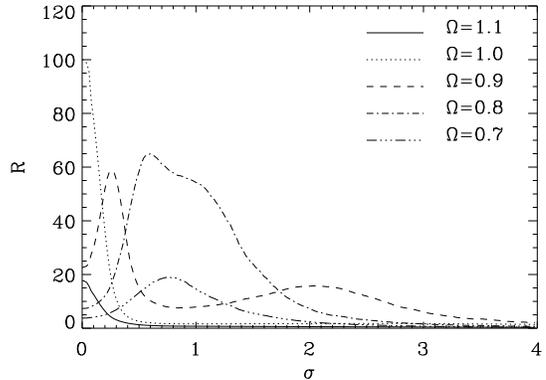}
    \end{center}
\caption{\label{fig:1} Plot of the response $R$, Eq.(\ref{eq:R}), versus the diversity parameter $\sigma$ for different values of the frequency $\Omega$ of the external forcing. The natural frequencies $\omega_k$ of the oscillators have been drawn from a Gaussian distribution of mean $\omega_0=1$ and variance $\sigma^2$. Other parameters are: $m_k=1$, $\gamma_k=0.1$, $\kappa=1$.}
\end{figure}
\begin{figure}[h]
    \begin{center}
    \includegraphics[width=8cm]{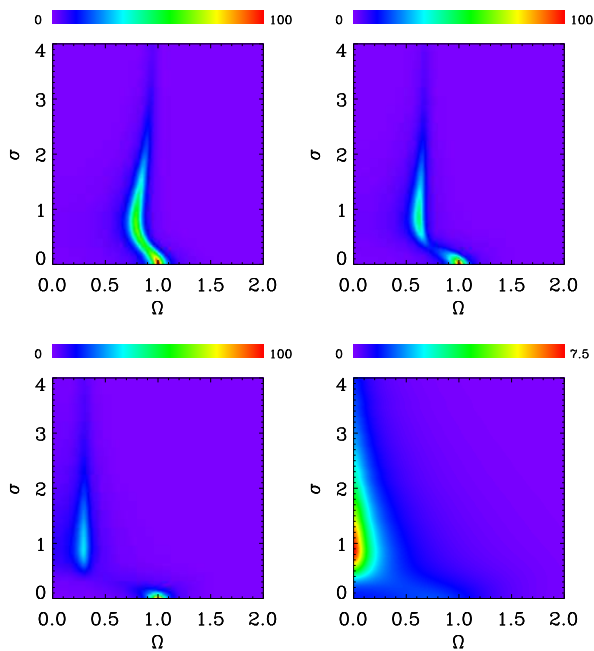}
\end{center}
\caption{\label{fig:2} Two-dimensional plot of the response $R$ versus the frequency of the external forcing $\Omega$ and the rms $\sigma$ of the Gaussian distribution (mean value $\omega_0=1$) of the internal frequencies $\omega_k$. Other parameters are $m_k=1$ and: $\gamma_k=0.1$, $\kappa=1.0$ (top left); $\gamma=0.1$, $\kappa=0.5$ (top right); $\gamma=0.1$, $\kappa=0.1$ (bottom left); $\gamma=1.0$, $\kappa=0.1$ (bottom right).}
\end{figure}

The main result of this paper is summarized in figure \ref{fig:1} where we plot the global system's response $R$, as given by Eq.(\ref{eq:R}), versus the diversity parameter $\sigma$ for different values of the frequency $\Omega$ of the external forcing. It is apparent from this figure that, for values of $\Omega$ smaller than the average value $\omega_0$, the response increases and goes through a maximum -a resonance- as the diversity parameter increases. In fact, as can be seen in the figure, for $\Omega$ close to $\omega_0$,  there can be two peaks in the resonance. Those peaks merge into a single one, with decreasing height,  as $\Omega$ decreases.  The regions of existence of the resonance can be better observed in a two dimensional plot of the response as a function of the diversity $\sigma$ and the frequency $\Omega$, see figure \ref{fig:2}. Note that, according to this figure, the resonant frequency moves with $\sigma$ to regions where there was no strong response in the absence of disorder.

The {\sl microscopic} argument to explain the resonance is very simple: When $\sigma=0$ all oscillators have the same frequency $\omega_0$. The response is  optimal for $\omega_0= \Omega$. If  $\omega_0\ne\Omega$ the oscillators are not at the optimal value. Then, as $\sigma$ increases, some oscillators will have frequencies close to $\Omega$ and will tend to have a larger response. It is those oscillators that, through the coupling term, pull the others and produce the large response of the collective variable $\bar x$.  

\begin{figure}[h]
    \begin{center}
    \includegraphics[width=8cm]{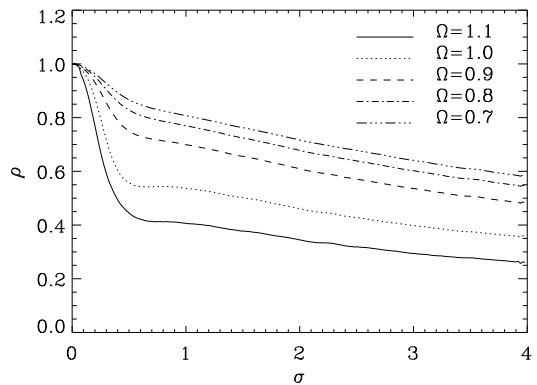}
    \end{center}
\caption{\label{fig:3} Plot of the measure of order $\rho$, Eq.(\ref{eq:rho}), characterizing the dispersion of the position of the oscillators, as a function of the diversity parameter $\sigma$ for different values of the frequency $\Omega$ of the external forcing. The natural frequencies $\omega_k$ of the oscillators have been drawn from a Gaussian distribution of mean $\omega_0=1$ and variance $\sigma^2$. Other parameters are: $m_k=1$, $\gamma_k=0.1$, $\kappa=1$.}

\end{figure}

\begin{figure}[h]
    \begin{center}
    \includegraphics[width=8cm]{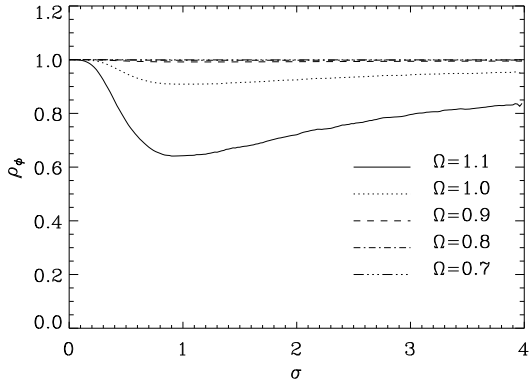}
    \end{center}
\caption{\label{fig:4} Same as Fig.\ref{fig:3} for the Kuramoto order parameter $\rho_{\phi}$, Eq(\ref{eq:rhophi}), characterizing the dispersion in phase of the oscillators.}
\end{figure}

\begin{figure}[h]
    \begin{center}
    \includegraphics[width=8cm]{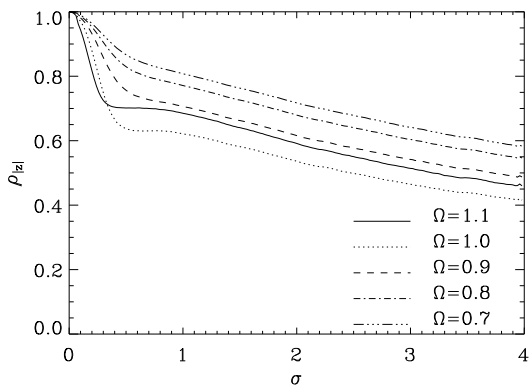}
    \end{center}
\caption{\label{fig:5} Same as Fig.\ref{fig:3} for $\rho_{|z|}$, Eq(\ref{eq:rhophi}), characterizing the dispersion in modulus of the oscillators.}
\end{figure}

\begin{figure}[h]
    \begin{center}
    \includegraphics[width=8cm]{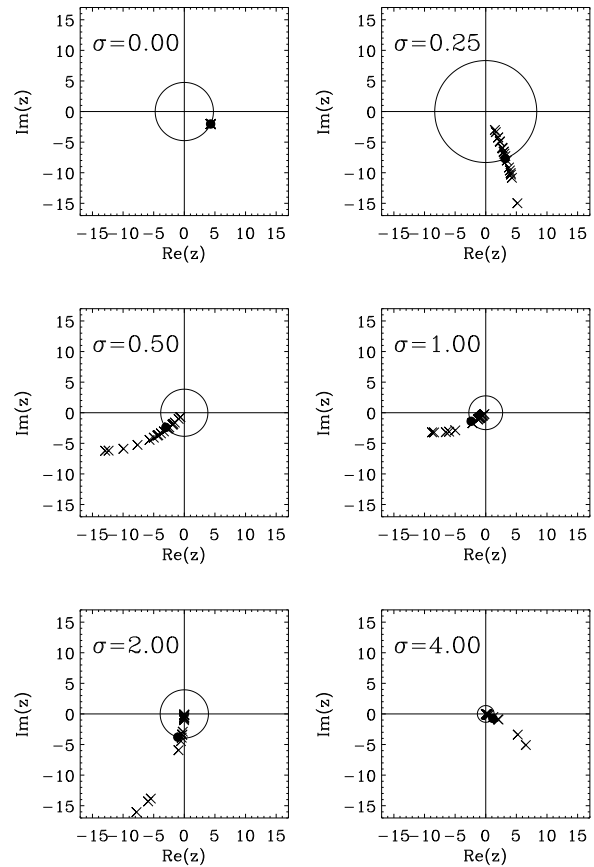}
    \end{center}
\caption{\label{fig:6} Phase space $z_k=(x_k,\Omega^{-1}\dot x_k)$ for different values of the diversity parameter $\sigma$. The natural frequencies $\omega_k$ of the oscillators have been drawn from a Gaussian distribution of mean $\omega_0=1$ and variance $\sigma^2$. Other parameters are: $m_k=1$, $\gamma_k=0.1$, $\kappa=1$, $\Omega=0.9$.  The solid dot is the position of the mean value $\bar z$ of $N=10000$ oscillators, while the crosses indicate the position of $20$ of those oscillators. The circle has the radius of the modulus of $\bar z$. All points rotate clockwise with the circular frequency $\Omega$ of the external forcing.}
\end{figure}

As argued in \cite{TMTG:2006}, the increase in the response is accompanied by a decrease in the order (but not necessarily vice-versa). This is clearly illustrated in figure \ref{fig:3} where we plot the measure of order $\rho$, as defined in Eq.(\ref{eq:rho}),  as a function of the diversity parameter $\sigma$ for different values of the frequency $\Omega$ of the external forcing. As shown in figure \ref{fig:4}, the Kuramoto order parameter stays close to the value $\rho_{\phi}=1$ for values of $\sigma$ where the resonance appears but the dispersion in the amplitude, as measured by $\rho_{|z|}$, increases with $\sigma$, see figure \ref{fig:5}. This shows that the oscillators lose their order through a dispersion of the modulus of their amplitude. In other words, they follow the external signal with similar phases, but with different amplitudes. Figure \ref{fig:6} helps to demonstrate this effect. We plot in this figure the trajectories in the complex plane $z$, equivalent to the phase space $(x,\dot x)$. When $\sigma=0$, all units have the same amplitude and phase and oscillate with the external frequency $\Omega$. As $\sigma$ increases, the average value $\bar z$ oscillates with a larger amplitude (and always with the frequency $\Omega$ of the external forcing) and the individual units oscillate with nearly the same phase but with different amplitudes, producing the dispersion that can be seen in the figure. 

\begin{figure}[h]
    \begin{center}
    \includegraphics[width=8cm]{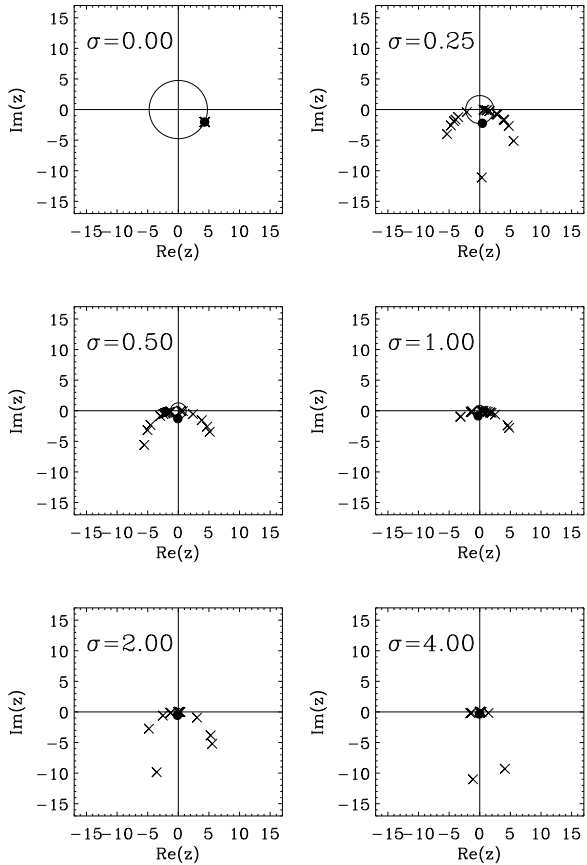}
    \end{center}
\caption{\label{fig:7} Same as figure \ref{fig:6} in the no coupling case,  $\kappa=0$. }
\end{figure}

\begin{figure}[h]
    \begin{center}
    \includegraphics[width=8cm]{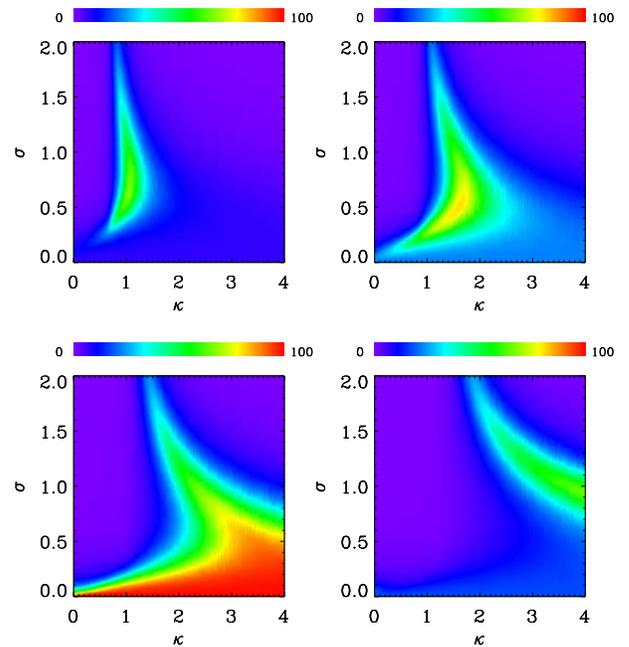}
\end{center}
\caption{\label{fig:8} Two-dimensional plot of the response $R$ versus the coupling $\kappa$ and the rms $\sigma$ of the Gaussian distribution (mean value $\omega_0=1$) of the internal frequencies $\omega_k$. Other parameters are $m_k=1$, $\gamma_k=0.1$ and: $\Omega=0.8$ (top left), $\Omega=0.9$ (top right), $\Omega=1.0$ (bottom left), $\Omega=1.1$ (bottom right).}
\end{figure}

\begin{figure}[h]
    \begin{center}
    \includegraphics[width=8cm]{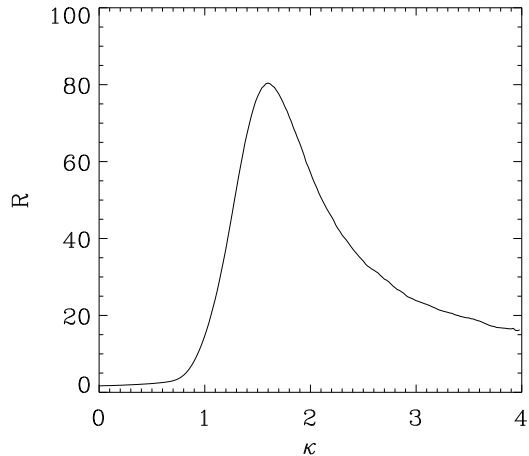}
    \end{center}
\caption{\label{fig:9} Plot of the response $R$, Eq.(\ref{eq:R}), versus the coupling parameter $\kappa$. The natural frequencies $\omega_k$ of the oscillators have been drawn from a Gaussian distribution of mean $\omega_0=1$ and variance $\sigma^2$. Other parameters are: $m_k=1$, $\gamma_k=0.1$,  $\Omega=0.9$, $\sigma=0.5$.}
\end{figure}

Coupling is essential for the resonance effect. In figure \ref{fig:7} we plot the phase space in the case of absence of coupling for the same set of parameters as those of figure \ref{fig:6}. As can be seen in this figure, the response always decreases with increasing the diversity. What it is somewhat surprising is that a too large coupling also diminishes the response. This is observed in the surface plot of figure \ref{fig:8}. The decrease  of the response for large coupling is also evident in figure \ref{fig:9} where we use the same value of the frequency as in figures \ref{fig:6} and \ref{fig:7}.

\begin{figure}[h]
    \begin{center}
    \includegraphics[width=8cm]{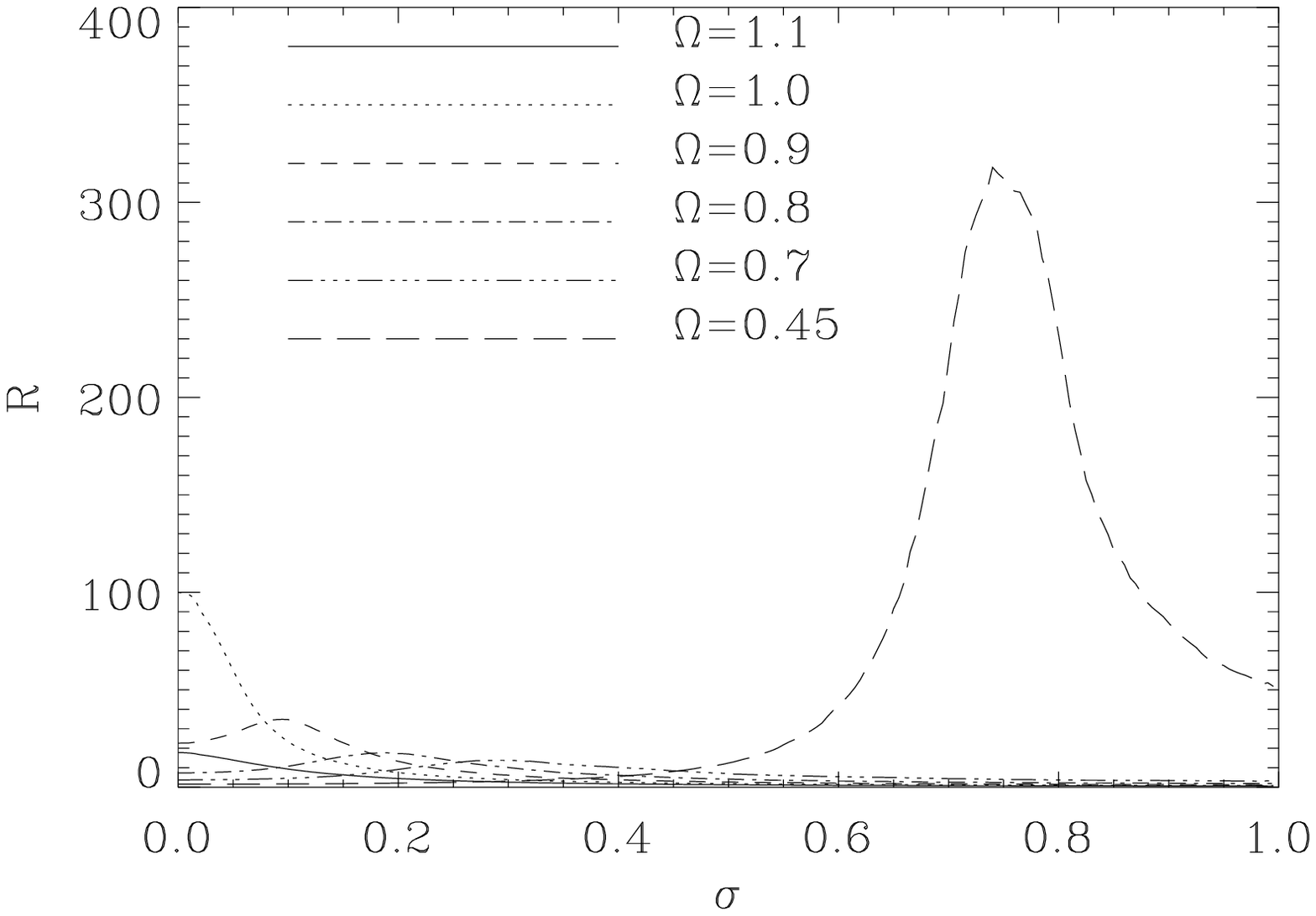}
    \end{center}
\caption{\label{fig:10} Plot of the response $R$, Eq.(\ref{eq:R}), versus the diversity parameter $\sigma$ for different values of the frequency $\Omega$ of the external forcing. The natural frequencies $\omega_k$ and the damping parameters $\gamma_k$ have been drawn from a common lognormal distribution of mean values $\omega_0=1$, $\gamma_0=0.1$, and rms $\sigma\omega_0$,  $\sigma\gamma_0$, respectively (see the text for details). Other parameters are:  $m_k=1$, $\kappa=0.2$.}
\end{figure}

So far, we have focused our attention on the case of variability in the distribution of internal frequencies $\omega_k$. It is not the intention of this paper to overwhelm the reader with many figures covering the many possibilities of dispersion in the different parameters. We want to finish by plotting in the final figure \ref{fig:10} the resonance effect that appears when both the frequencies $\omega_k$ and the damping coefficients $\gamma_k$ vary from one unit to the other. To maintain positivity of the damping coefficients, we have chosen a log-normal distribution for both parameters. More concretely, we have taken $\omega_k=\omega_0\tau_k$ and $\gamma_k=\gamma_0\tau_k$, where $\tau_k$ follows a log-normal distribution of mean $1$ and rms $\sigma$. It is clear that a strong resonance effect appears as a function of the diversity $\sigma$ (measured now as the relative, normalized to the mean value, rms of the log-normal distributions). Note that, for this particular set of parameter values, the maximal response is much larger than the maximum response obtained when the external frequency equals the mean value $\omega_0=\Omega$.

\section{Conclusion}
\label{conclusions}

In this paper we have studied an ensemble of globally coupled linear oscillators subjected to periodic forcing. Diversity appears as each unit of the ensemble has different characteristic parameters, although we have focused our main attention on the case in which only the natural frequencies  vary from one oscillator to another. Our intention has been to analyse with some detail an exactly solvable model that displays the phenomenon of diversity-induced resonance, namely, that the global response to the external forcing is enhanced by the presence of diversity in the units. We have been able to give explicit expressions for the global response and some measures of the order in the positions, phases and modulus of the units of the ensemble.

The details of the phenomenon bear some similarities, but also some differences, with previous studies in more complicated systems. The main microscopic mechanism is as follows: In the no-diversity case,  all units have the same natural frequency, which is different from the frequency of the external forcing. As the diversity in frequencies increases, a fraction of the units have a frequency that matches that of the external forcing and are then resonant with it. Those units are able, through the coupling terms, to pull the other units and the global, average, variable is able to display large amplitude oscillations. As the diversity increases even further, the difference between the units increases. Some of them have then a frequency too far away from the external frequency and tend to follow the forcing with a small amplitude, hence diminishing the global response. The main difference with the mechanism analyzed in double-well systems\cite{TMTG:2006} is that, as described in the introduction, in those systems the units that are able to follow the forcing vary from time to time.  That model is thus more useful to model situations of a changing environment.

The analytical solution of the linear equations of motion allows a full analysis of the problem. As implied in reference \cite{TMTG:2006}, see also \cite{TTV:2007}, the enhancement in the response appears together with a decrease in the macroscopic order. We have introduced appropriate  measures of this order in terms of the dispersion in the locations of the oscillators in phase space. Those measures show that the loss of order is mainly in the amplitude of the oscillations whereas the phases show a much smaller dispersion. We are currently interested in studying which of these features remain when considering the enhancement in the response of non-linear ocillators induced by diversity. The results wil be reported elsewhere.

\section*{Acknowledgments}

We acknowledge financial support from the EU NoE BioSim, LSHB-CT-2004-005137, and project FIS2007-60327 from MEC (Spain).  One of us (JDG) acknowledges the support of grants from the  G. Harold and Leila Y. Mathers Foundation and NSF grant number DMR0702890.

\end{document}